\documentclass[a4paper,fleqn]{cas-dc}

\usepackage[numbers]{natbib}
\usepackage{amsmath}
\usepackage{color}
\usepackage{algorithmic}
\usepackage{array}

\usepackage{fixltx2e}
\usepackage{stfloats}
\usepackage{float}
\usepackage{textcomp}
\usepackage{multirow}

\usepackage[font=small,labelfont=bf,labelsep=none]{caption}
\def\tsc#1{\csdef{#1}{\textsc{\lowercase{#1}}\xspace}}
\tsc{WGM}
\tsc{QE}
\tsc{EP}
\tsc{PMS}
\tsc{BEC}
\tsc{DE}

\begin{document}
\let\WriteBookmarks\relax
\def\floatpagepagefraction{1}
\def\textpagefraction{.001}
\shorttitle{Does Non-COVID19 Lung Lesion Help? Investigating Transferability in COVID-19 CT Image Segmentation}
\shortauthors{Yixin Wang et~al.}

\title [mode = title]{Does Non-COVID19 Lung Lesion Help? Investigating Transferability in COVID-19 CT Image Segmentation}                      

\credit{Conceptualization of this study, Methodology, Software}

\address[1]{Institute of Computing Technology, Chinese Academy of Sciences, University of Chinese Academy of Sciences, Beijing, China}

\author[1,2]{Yixin Wang}[orcid=0000-0002-8062-0765]
\author[1,2]{Yao Zhang}
\author[1,2]{Yang Liu}

\credit{Data curation, Writing - Original draft preparation}

\address[2]{AI Lab, Lenovo Research, Beijing, China}

\author[2]{Jiang Tian}
\author[2]{Cheng Zhong}
\author[2]{Zhongchao Shi}
\address[3]{Lenovo Corporate Research \& Development, Lenovo Ltd., Beijing, China }
\author[1,3]
{Yang Zhang}
\cormark[1]

\author[1,3]
{Zhiqiang He}
\cormark[1]

\cortext[cor1]{Corresponding author}
\nonumnote{E-mail addresses: wangyixin19@mails.ucas.ac.cn (Yixin Wang), zhangyang20@lenovo.com (Yang Zhang), hezq@lenovo.com (Zhiqiang He)}

\begin{abstract}
\textbf{Background and Objective}: 
Coronavirus disease 2019 (COVID-19) is a highly contagious virus spreading all around the world. Deep learning has been adopted as an effective technique to aid COVID-19 detection and segmentation from computed tomography (CT) images. The major challenge lies in the inadequate public COVID-19 datasets. 
Recently, transfer learning has become a widely used technique that leverages the knowledge gained while solving one problem and applying it to a different but related problem. However, it remains unclear whether various non-COVID19 lung lesions could contribute to segmenting COVID-19 infection areas and how to better conduct this transfer procedure. This paper provides a way to understand the transferability of non-COVID19 lung lesions and  a better strategy to train a robust deep learning model for COVID-19 infection segmentation.\\
\textbf{Methods}: 
Based on a publicly available COVID-19 CT dataset and three public non-COVID19 datasets, we evaluate four transfer learning methods using 3D U-Net as a standard encoder-decoder method. i) We introduce the multi-task learning method to get a multi-lesion pre-trained model for COVID-19 infection. ii) We propose and compare four transfer learning strategies with various performance gains and training time costs. Our proposed Hybrid-encoder Learning strategy introduces a Dedicated-encoder and an Adapted-encoder to extract COVID-19 infection features and general lung lesion features, respectively. An attention-based Selective Fusion unit is designed for dynamic feature selection and aggregation.\\
\textbf{Results}: 
Experiments show that trained with limited data, proposed Hybrid-encoder strategy based on multi-lesion pre-trained model achieves a mean DSC, NSD, Sensitivity, F1-score and Accuracy of 0.704, 0.735, 0.682, 0.707 and 0.994, respectively, with better genetalization and lower over-fitting risks for segmenting COVID-19 infection.\\
\textbf{Conclusions}: 
The results reveal the benefits of transferring knowledge from non-COVID19 lung lesions, and learning from multiple lung lesion datasets can extract more general features, leading to accurate and robust pre-trained models. We further show the capability of the encoder to learn feature representations of lung lesions, which improves segmentation accuracy and facilitates training convergence. In addition, our proposed Hybrid-encoder learning method incorporates transferred lung lesion features from non-COVID19 datasets effectively and achieves significant improvement.
These findings promote new insights into transfer learning for COVID-19 CT image segmentation, which can also be further generalized to other medical tasks.
\end{abstract}




\maketitle

\section{Introduction}
%
%
%
%
In December 2019, the coronavirus disease 2019 (COVID-19) broke out and has become a global challenge since then. This new virus was spreading rapidly, affecting countries worldwide. Up to 30 July 2020, 16,341,920 identified cases of COVID-19 have been reported in over 216 countries and territories, resulting in 650,805 deaths. This severe disease has been declared as a Public Health Emergency of International Concern by the World Health Organization (WHO).

A gold standard method to screen the COVID-19 patients is the real-time reverse transcription-polymerase chain reaction(rRT-PCR)\cite{Chane}, which reports the results within few hours to 2 days and requires repeated tests\cite{coconet,PEREIRA}. 
Researchers found that chest computed tomography (CT) images show strong ability to capture ground glass and bilateral patchy shadows which are typical CT features in affected patients\cite{Chung}. Thus, chest CT can help identify distinguishing patterns and features in patients. Given that traditional CT imaging analysis methods are time-consuming and laborious, it is of great importance to develop artificial intelligence (AI) systems to aid COVID-19 diagnosis\cite{Jin}. 

Segmentation in CT slices is a key component of the diagnostic work-up for patients with COVID-19 infection in clinical practice\cite{review}. Recently, deep learning with CNNs has showed significant performance improvements on the automatic detection and automatic extraction of essential features from CT images, related to the diagnosis of the Covid-19. Though deep learning has made great progress in medical image segmentation, it remains a challenging task in the field of COVID-19 lung infection, as existing public datasets on COVID-19 are relatively small and weakly labelled. Thus, training deep networks from scratch with inadequate data and task-specific nature such as COVID-19 infection may lead to over-fitting and poor generalization.

Transfer learning is an effective method to solve this problem, which helps to leverage knowledge and latent features from other datasets and avoids over-parameterization. In transfer learning, successful deep learning models such as ResNet, DenseNet and GoogLeNet have been trained on large datasets such as ImageNet. These pre-trained models have proven impressive performance on natural image downstream tasks. However, there exists no large-scale annotated medical image datasets as data acquisition is difficult, and high-quality annotations are expensive. Recent research\cite{transfusion} shows that transfer learning from natural image datasets to medical tasks produces very limited performance gain. In particular, though with large medical datasets for pre-training, transfer learning in the task of COVID-19 infection segmentation is still much more difficult:
1) The shape, texture and position of COVID-19 infections are in high variation.
2) Existing large medical CT datasets differ in domains with COVID-19 datasets. Thus, similar in domain, a pre-trained model from lung lesions may share more knowledge with COVID-19 infection and learn some general-purpose visual representations for lung lesions. 

Evidence shows that larger datasets are necessarily better for pre-training and the diversity of datasets is extremely important\cite{Not-so-supervised}. In medical domain, pre-training from medical datasets, especially chest CT datasets tends to be more homogeneous compared to non-medical and other medical areas' data. Thus, non-COVID19 lung lesion CT imaging manifestations may serve as potential profit for COVID-19 segmentation. Existing works have proven that multi-task training through simply fusing different lesions with COVID-19 affects model's representation ability for COVID-19 infection segmentation\cite{COVID-19-SegBenchmark}. Therefore, with limited COVID-19 datasets, 1) whether these non-COVID19 lesions help and to what extent they can contribute to COVID-19 infection segmentation? 2) How to train a better pre-trained model using these non-COVID19 datasets for transfer learning? 3) In what manners can the pre-trained models fitted on non-COVID19 lesions be effectively transferred to COVID-19?

In this paper, we aim to answer the above questions which are significant for COVID-19 segmentation. To our best knowledge, this is the first study to explore the transferability of non-COVID19 datasets for COVID-19 CT images segmentation.

Our contributions are as follows.
\begin{itemize}
    \item We experimentally assess the extent of contributions from non-COVID19 to COVID-19 infection segmentation. We found that despite the disparity between non-COVID19 lung lesion images and COVID-19 infection images, pre-training on the large scale well-annotated lung lesions may still be transferred to benefit COVID-19 recognition and segmentation.
\end{itemize}  
\begin{itemize}
    \item We conduct extensive experiments using various non-COVID19 lung lesion datasets. 
    Although pre-training on a single non-COVID19 dataset is unstable among different transfer strategies, learning from different non-COVID19 lesions demonstrates promising performance since such a multi-task learning process can share knowledge from different related tasks and discover common and general representations of lung lesions.
\end{itemize}
\begin{itemize}
    \item We design four different transfer learning strategies with various performance gains and training time costs: Continual Learning, Body Fine-tuning, Pre-trained Lesion Representations and Hybrid-encoder Learning. The Hybrid-encoder strategy effectively combines both non-COVID19 and COVID-19 features and shows the best performance. We further conclude that it is possible to freeze the encoder and train only the decoder on COVID-19 tasks during fine-tuning, which provides significant performance gains and fast convergence speed.
\end{itemize}
\section{Background}
In this section, we review the recent related research on COVID-19 CT images, COVID-19 segmentation and transfer learning for COVID-19.
\subsection{Research on COVID-19 CT Images}
CT is widely used in the screening and diagnosis of COVID-19 since ground-glass opacities and bilateral patchy shadows are the most relative imaging features in pneumonia associated with infections. Recent research on CT-based diagnosis for COVID-19 has indicated great performance. Compared with traditional CT images processing, artificial intelligence (AI) serves as a core technology, enabling a more accurate and efficient solution.
The machine learning-based CT radiomics models such as random forest, logistic regression showed feasibility and accuracy for predicting hospital stay in patients affected in the work of \cite{Qi}. 
Machine learning method was also adopted by Tang \emph{et al.}\cite{Tang} to realize automatic severity assessment (non-severe or severe) of COVID-19 based on chest CT images, and to explore the severity-related features from the resulting assessment model.
Gozes \emph{et al.}\cite{O.Gozes} presented a system that utilized robust 2D and 3D deep learning models, modifying and adapting existing AI models and combining them with clinical understanding.  Huang \emph{et al.}\cite{huang} monitored the disease progression and understood the temporal evolution of COVID-19 quantitatively using serial CT scan by an automated deep learning method.

\subsection{Research on COVID-19  Segmentation}
Segmentation is an essential step in AI-based image processing and analysis\cite{review}. In particular, segmenting the regions of interests (ROIs) of COVID-19 infections is crucial and helpful for doctors to make further assessment and quantification. However, manual contouring of these infections is time-consuming and tedious. Although plenty of methods have been explored on COVID-19 diagnosis and classification, there are very few works on the segmentation of COVID-19 infection due to its great annotation challenges. 

Shan \emph{et al.}\cite{shan} developed a DL-based system for automatic segmentation and quantification of infection regions and adopted a human-in-the-loop (HITL) strategy to accelerate the manual delineation of CT training.
Zheng \emph{et al.}\cite{Zheng} designed a  weakly-supervised deep learning algorithm to investigate the potential of a deep learning-based model for automatic COVID-19 detection on chest CT volumes using the weak patient-level label.
Based on semi-supervised learning, Fan \emph{et al.}\cite{infnet} presented a COVID-19 Lung Infection Segmentation Deep Network (Inf-Net) for CT slices based on randomly selected propagation.
Yan \emph{et al.}\cite{Yan} introduced a feature variation block which adaptively  adjusted the global properties of the features for segmenting COVID-19 infection.

\subsection{Research on COVID-19 Transfer Learning}
In transfer learning, deep models are first trained on large datasets such as ImageNet, then these pre-trained models are fine-tuned on different downstream tasks. Several studies \cite{Not-so-supervised}, \cite{Transfer_med2}, \cite{Transfer_med3} have investigated transfer learning methodologies on deep neural networks applied to medical image analysis tasks.  Plenty of works used networks pre-trained on natural images to extract features and followed by another classifier\cite{Cheplygina_2018},\cite{Carneiro}. Carneiro \emph{et al.}\cite{Carneiro} replaced the fully connected layers of a pre-trained model with a new logistic layer and only trained the appended layer, yielding promising results for classification of unregistered multi-view mammograms. Other studies performed layer fine-tuning on the pre-trained networks for adapting the learned features to the target domain. In \cite{Bar}, CNNs were pre-trained as a feature generator for chest pathology identification. Gao \emph{et al.}\cite{Gao} fine-tuned all the layers of a pre-trained CNN to classify interstitial lung diseases. Ghafoorian \emph{et al.}\cite{Ghafoorian_2017} trained a CNN on legacy MR images of brain and evaluated the performance of the domain-adapted network on the same task with images from a different domain. 

Due to the limited labeled COVID-19 data, several transfer learning methods have been applied to address this problem. Chouhan \emph{et al.}\cite{Chouhan} proposed an ensemble approach of transfer learning using pre-trained models trained on ImageNet. Researchers used five different pre-trained models and analyzed their performance in chest X-ray images. In the work of \cite{transfer2}, several deep CNNs were employed for automatic COVID-19 infection detection from X-ray images through tuning parameters. 
Majeed \emph{et al.}\cite{transfer4} presented a critical analysis for 12 off-the-shelf CNN models and proposed a simple CNN architecture with a small number of parameters that performed well on distinguishing COVID-19 from normal X-rays.
By combining three different models which were fine-tuned on 3 datasets, Misra \emph{et al.}\cite{transfer6} designed a multi-channel ensemble TL method based 
on ResNet-18 in such a way that the model could extract more relevant features for each class and identify COVID-19 features more accurately from the X-ray images. 

Even though the above studies on transfer learning present inspiring achievement for COVID-19 research, there are several limitations: 1) They only focus on the ensemble of existing CNNs, but ignore the contribution of various datasets for pre-training. 2) Their studies are limited to X-ray dataset and only dedicate to COVID-19 detection and classification. 3) They lack an in-depth study on the transferability of different transfer manners related to COVID-19 infections. Our work contributes to a much more difficult task of semantic segmentation in COVID-19 CT images. We explore the transferability from the perspective of transferring knowledge from various non-COVID19 lung lesions. Moreover, we investigate better transfer methods to assist COVID-19 segmentation. 

\section{Methodology}
In this section, we briefly describe our backbone network in \ref{Sec 3.1}, then introduce the multi-task learning method to get a multi-lesion pre-trained model in \ref{Sec 3.2}. We further give detailed illustration on four transfer strategies employed in our work in \ref{Sec 3.3}-\ref{Sec 3.5}.
\subsection{Encoder-Decoder Network}
\label{Sec 3.1}
The U-Net is a commonly used network for medical semantic segmentation. As an advanced architecture, it has a U-shape like structure with an encoding and a decoding signal path. The encoder serves as a contraction to capture semantically image contextual features. The decoder is a symmetric expanding path recovering spatial information. The two paths are connected using skip connections on each same level, which recombine with essential high-resolution features from the encoding path. In this work, we make some minor changes following nnU-Net\cite{nnunet}. Original batch normalization and ReLU are replaced by instance normalization and leaky ReLU. What's more, deep supervision loss\cite{dou} is aggregated to obtain multi-level deep supervision and facilitate the training process. 
\subsection{Pre-trained Multi-lesion Learning}
\label{Sec 3.2}
Multi-task learning is an effective method to share knowledge among different related tasks. As for segmentation tasks, the performance of each task highly depends on the similarity among these tasks. Due to the large domain distance among those existing non-COVID19 lung lesion datasets, fusing these lesions to train a multi-task segmentation model tends to underperform on each single task. However, this multi-task training can exploit the shared knowledge which is essential for learning some general-purpose visual representations about lung lesions. Therefore, in our work, besides separately training segmentation models on each non-COVID19 dataset as pre-trained models for transfer learning, we provide a multi-lesion model learning from multiple lung lesions. Compared with learning from separate tasks, this multi-task strategy leads to a more robust pre-trained model across all lesion tasks and empowers downstream COVID-19 task.
\subsection{Continual Learning}
\label{Sec 3.3}
Continual Learning aims to learn from an endless stream of tasks\cite{houlsby}. This paradigm is capable of learning consecutive tasks without forgetting how to perform previously trained tasks. This is challenging that the training process tends to lose knowledge from the previous tasks due to the information relevant to current tasks. To avoid this, we adopt a training schedule to pre-train the model. The value of the initial learning rate is set as 0.01 and decays throughout the training process following the `poly' learning rate policy $(1-\text { epoch } / \text { epoch }_{\text {max }})^{0.9}$. In this way, the learning rate is decreasing continuously so that the weight parameters after training non-COVID19 lesion tasks tend to follow its training process while slightly being updated by the current COVID-19 infection task. 
\subsection{Fine-tuning}
\label{Sec 3.4}
Fine-tuning is the most standard strategy to transfer knowledge from domains where data are abundant. In general, it is conducted by copying the weights from a pre-trained network and tuning them on the downstream task. Recent work\cite{sun} shows that fine-tuning can enjoy better performance on small datasets.
It is of additional interest to assess the contribution of the encoder and decoder relative to learning COVID-19 knowledge. While the progressive reinitializatons demonstrate the incremental effect of each layers, it is unnecessary to prob the extent of localized reinitialization because our encoder-decoder network is not very deep. Therefore, we adopt two strategies for tuning a non-COVID19 lesion model on the COVID-19 downstream task. \\
\subsubsection{Body Fine-tuning}
 Due to the large domain difference between pre-trained tasks and downstream tasks, the most secured fine-tuning method is Body Fine-tuning, which means all parameters of the pre-trained models are used as the initial values to complete the training process of the model. When we train COVID-19 infection networks, all of the parameters are assigned initially from the pre-trained models on non-COVID19 lesions. In this fine-tuning strategy, the update of parameters largely depends on COVID-19 infection training process itself. Thus, it is a conservative fine-tuning approach that the training task of COVID-19 infection is not affected by the upstream pre-trained non-COVID19 models too much.
\\
\subsubsection{Pre-trained Lesion Representations}
Our segmentation model is an encoder-decoder architecture and the encoder serves as a series of convolution operations to encode image features into context representations. These representations are trained on large relative datasets from upstream tasks, and fed as features to downstream ones. In natural language processing tasks, features extracted from internal representations of sequence language models are encoded as pre-trained text representations\cite{houlsby}, \cite{sun}. In this fine-tuning strategy, we aim to learn general lung lesion features, which we call Pre-trained Lesion Representations. We first train models on non-COVID19 lung lesion datasets. The encoders of these models are capable of encoding lesion features. In other words, the encoders' parameters can exhibit some transferability. Thus, in the following fine-tuning process on COVID-19 dataset, we preserve and freeze them while only fine-tuning and re-training the decoding parts.
\subsection{Hybrid-encoder Learning}
\label{Sec 3.5}
\begin{figure}[!t]
\centering
\includegraphics[width=3.5in]{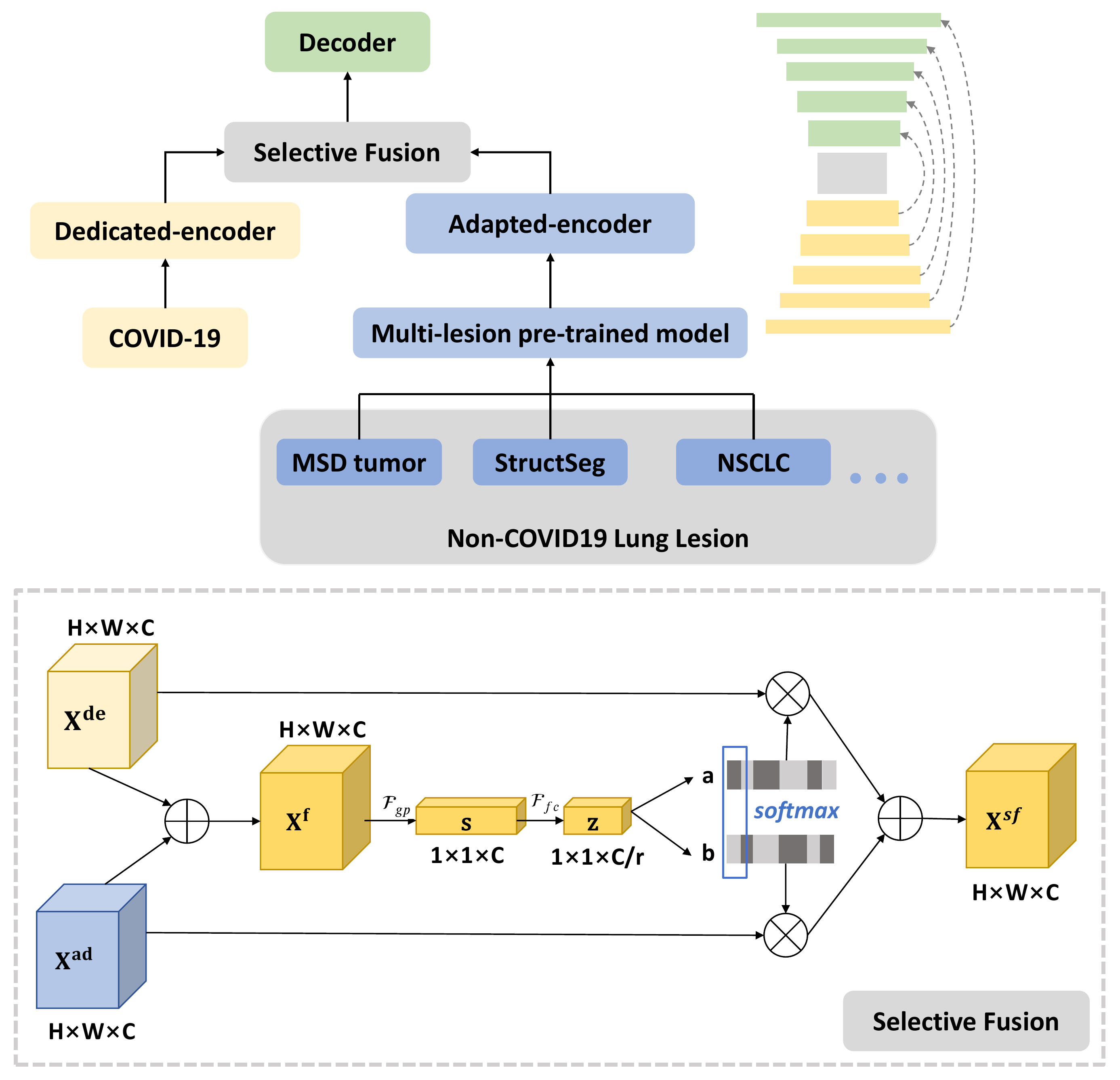}
\caption{An overview of Hybrid-encoder transfer learning strategy. The U-Net network consists of a parallel encoder and a shared decoder. Features from Dedicated-encoder and Adapted-encoder are aggregated via a Selective Fusion unit. For simplicity the figure shows 2D images.}
\label{method}
\end{figure}

During performing the above fine-tuning strategies, we face two challenges: 1) Body Fine-tuning easily falls into over-fitting because the downstream COVID-19 infection dataset is much smaller. 2) Utilizing the encoder of pre-trained models to capture feature representation is unstable, because the label spaces and losses for the upstream and downstream tasks differ inevitably. For example, though they are all lung lesions, they differ in appearance and shape. Therefore, we present a new transfer learning strategy for COVID-19, which incorporates three key properties:
\begin{itemize}
    \item It leverages the transferred knowledge from non-COVID19 lung lesions.
\end{itemize}
\begin{itemize}
    \item It gains stable performance improvement in both training from scratch and transfer learning methods.
\end{itemize}
\begin{itemize}
    \item It shows no obvious training time increase.
\end{itemize}
To achieve these properties, we propose a Hybrid-encoder architecture. As shown in Fig.\ref{method}, we enhance the standard U-Net network by equipping with two encoders with the same architecture: Dedicated-encoder and Adapted-encoder. Furthermore, an attention-based Selective Fusion unit is developed to aggregate information from both of the encoders by determining two sets of learnable weights.
\subsubsection{Dedicated-encoder}
The Dedicated-encoder is a task-specific feature extractor, focusing on segmenting COVID-19 infection through reinitializing all the parameters. These parameters $\Theta_{de}$ are all COVID19-specific which enable continuing update to the network. 
\subsubsection{Adapted-encoder}
The Adapted-encoder is an auxiliary feature extractor, aiming to learn general lung lesion features. Based on Pre-trained Multi-lesion learning in \ref{Sec 3.2}, we pre-train our 3D U-Net network's encoder to obtain dense representations of general lung lesions. When being transferred to target COVID-19 task, this pre-trained encoder serves as an adapted-encoder by totally freezing its pre-trained parameters as $\Theta_{ad}$. 
 
Given a COVID-19 input volume as the $s^{th}$ sample $\mathrm{X_s}\in \mathbb{R}^{H \times W\times D}$, it is passed through the above paralleled encoders. Each encoder follows the same 3D U-Net, which consists of a number of stacked convolution layers and pooling layers. More specifically, let $x_i^{de}, x_i^{ad}$ be the output of $i^{th}$ layer of Dedicated-encoder and Adapted-encoder, respectively. These vectors can be obtained from the output of the previous layer $x_{i-1}^{de}$ and $x_{i-1}^{ad}$ by a mapping $\mathcal{H}(\cdot)$: 
\begin{equation}
\begin{aligned}
x_{i}^{de} =\mathcal{H}\left(x_{i-1}^{de}\right) = \sigma\left(\mathcal{I}\left(\mathbf{W^{de}} * x_{i-1}^{de}\right)\right), i \in\{0, \ldots, l\}, \\
x_{i}^{ad} =\mathcal{H}\left(x_{i-1}^{ad}\right) = \sigma\left(\mathcal{I}\left(\mathbf{W^{ad}} * x_{i-1}^{ad}\right)\right), i \in\{0, \ldots, l\},
\end{aligned} 
\end{equation}
where $W$ represents the weight matrix, $*$ denotes convoluton operation, $\mathcal{I}(\cdot)$ and $\sigma (\cdot)$ represent instance normalization and leaky ReLU, respectively. 

At the end of the encoding phase, both  Dedicated-encoder and Adapted-encoder representations with $C$ channels containing rich semantic information are  learned separately from COVID-19 and non-COVID19 datasets, denoted as $\mathrm{X}^{\mathrm{de}}=f_{de}\left(\mathrm{X}_{s}; \Theta_{de}\right)$ and $\mathrm{X}^{\mathrm{ad}}=f_{ad}\left(\mathrm{X}_{s}; \Theta_{ad}\right)$,
and the two encoders are parameterized by $\Theta_{de}$ and $\Theta_{ad}$, respectively. 
\subsubsection{Selective Fusion}
Inspired by \cite{sknet}, we design a Selective Fusion operation to combine and aggregate the information from both encoders to obtain a global and comprehensive representation for decoding phase.
To achieve this, we first fuse $\mathrm{X}^{\mathrm{de}}$ and $\mathrm{X}^{\mathrm{ad}}$ using element-wise summation operation $\oplus$ as follows:
\begin{equation}
\mathrm{X}^{\mathrm{f}}=\mathrm{X}^{\mathrm{de}} \oplus \mathrm{X}^{\mathrm{ad}}.
\end{equation}
We then apply global average pooling to shrink $\mathrm{X}^{\mathrm{f}}$ through its 3D spatial dimensions $H \times W \times D$. The $c^{th}$ element of channel-wise statistics $\mathbf{s} \in \mathbb{R}^{C}$ is calculated by:
\begin{equation}
s_{c}=\mathcal{F}_{g p}\left(\mathrm{X}^{\mathrm{f}}\right)=\frac{1}{H \times W \times D} \sum_{i=1}^{H} \sum_{j=1}^{W} \sum_{l=1}^{D}\mathrm{X}^{\mathrm{f}}(i, j, l).
\end{equation}
The output $s_{c}$ can be interpreted as integrated local descriptors to provide selection guidance. In order to exploit channel-wise dependencies, we conduct fully connection (fc) via $\mathbf{W} \in \mathbb{R}^{\frac{C}{r} \times C}$ to reduce the dimension to $\mathbf{z} \in \mathbb{R}^{\frac{C}{r} \times 1}$ with a reduction ratio $r$:
\begin{equation}
\mathbf{z}=\mathcal{F}_{f c}(\mathbf{s})=\sigma(\mathcal{I}(\mathbf{W} * \mathbf{s})).
\end{equation}
To adaptively select different information from two encoders, we utilize softmax operation to obtain soft-attention across channels by:
\begin{equation}
\mathbf{a}=\frac{e^{\mathbf{A} \mathbf{z}}}{e^{\mathbf{A} \mathbf{z}}+e^{\mathbf{B} \mathbf{z}}}, \mathbf{b}=\frac{e^{\mathbf{B} \mathbf{z}}}{e^{\mathbf{A} \mathbf{z}}+e^{\mathbf{B} \mathbf{z}}},
\end{equation}
where $\mathbf{A},\mathbf{B} \in \mathbb{R}^{\frac{C}{r} \times C}$ and $\mathbf{a},\mathbf{b}$ represent the soft attention vectors for $\mathrm{X}^{\mathrm{de}}$ and $\mathrm{X}^{\mathrm{ad}}$, respectively. Through applying these soft-attentions using Eqn. \ref{sf}, features from these two paralleled encoders are dynamic selected as incorporated COVID-19 feature representations $\mathbf{X_{sf}}$ and then fed into the decoding part.
\begin{equation}
\label{sf}
\begin{aligned}
\mathbf{X_{sf}} &=\mathcal{H}\left(\mathbf{a} \cdot \mathrm{X^{de}} \oplus \mathbf{b} \cdot  \mathrm{X^{ad}} \right)
\end{aligned}
\end{equation}

\section{Experiments}
In this section, we will describe in detail the datasets, experimental setup and results of our investigation.
\subsection{Dataset Introduction}
\subsubsection{COVID-19 Dataset}
This dataset is released by Coronacases Initiative and Radiopaedia\footnote{https://github.com/ieee8023/covid-chestxray-dataset}. It is a publicly available COVID-19 volume dataset which contains 20 COVID-19 CT scans. In the work of \cite{COVID-19-SegBenchmark}, \cite{COVID-19-CT-Seg-Dataset}, left lung, right lung, and infections are well-labelled by two radiologists and verified by an experienced radiologist. Thus, with over 1800 annotated slices, this dataset serves as a downstream COVID-19 infection segmentation dataset for transfer learning in our work.
\subsubsection{Non-COVID19 Lung Lesion Datasets}
In order to better explore the transferability from various non-COVID19 lung lesions to COVID-19 infections, the following relationships need to be satisfied among these datasets: 1) The size of different lesion datasets are similar. 2) The shape, size and location of different lesion areas are relatively distinguishing.
Therefore, in this paper, we introduce three public datasets.

\textbf{MSD Lung Tumor}
This dataset was used in a crowd-sourced challenge of generalized semantic segmentation algorithms called the Medical Segmentation Decathlon (MSD) held in MICCAI 2018\footnote{http://medicaldecathlon.com/}. This dataset includes patients with non-small cell lung cancer from Stanford University (Palo Alto, CA, USA) publicly available through TCIA and previously utilized to create a radiogenomic signature\cite{MSD1}, \cite{MSD2}, \cite{MSD3}. The tumor regions are denoted by an expert thoracic radiologist on a representative CT cross section using OsiriX\cite{MSD4}. 63 3D CT scans with corresponding tumor segmentation masks are utilized in our paper.

\textbf{StructSeg Lung Cancer}
StructSeg organ dataset is a collection of 3D organ CT scans along with their segmentation ground-truth from 2019 MICCAI challenge\footnote{https://structseg2019.grand-challenge.org}. This dataset contains two types of cancers, nasopharyngeal cancer and lung cancer. We adopt the gross target volume segmentation of lung cancer from 50 patients. Each CT scan is annotated by an experienced oncologist and verified by another one.

\textbf{NSCLC Pleural Effusion}
This dataset is developed from a Non-Small Cell Lung Cancer (NSCLC) cohort of 211 subjects\footnote{https://wiki.cancerimagingarchive.net/display/Public/NSCLC-Radiomics}. 78 cases with pleural effusion are selected with their segmentation masks.

\subsection{Experimental Settings}
All the experiments are implemented in Pytorch and trained on NVIDIA Tesla V100 32GB GPU. For fair comparison, we follow the settings on COVID-19 dataset benchmarks in \cite{COVID-19-SegBenchmark}. 

For the COVID-19 dataset, we use 5-fold cross validation based on a pre-defined dataset split file. Each fold contains 4 scans (20$\%$) for training and 16 (80$\%$) for testing. Training fewer data is more suitable for exploring the contribution of transfer learning. 

For non-COVID19 lung lesion datasets, we all randomly select 80$\%$ of the data for training and the rest of 20$\%$ for validation. Pre-trained models based on these non-COVID19 lesions are all trained from scratch with random initial parameters using 3D U-Net network. Due to the limited number of different lesion cases, we do data pre-processing following nnU-Net\cite{nnunet}. The input patch size is set as 50$\times$160$\times$192 and batch size as 2. Stochastic gradient descent optimizer with an initial learning rate of 0.01 and a nesterov momentum of 0.99 are used for non-COVID19 pre-training. Reduction ratio $r$ is set as 16, following \cite{sknet}. We adopt a summation of Dice loss and Cross entropy loss as loss function.

\subsection{Evaluation Metrics}
Diagnostic evaluation is often used in clinical practice for disease diagnosis, patient follow-up or efficacy monitoring. Whether the results of a certain diagnostic evaluation are true, reliable and practical, will largely determine a reasonable medical decision. In this work, we introduce five evaluation metrics for the exploration of transferability. 

Dice similarity coefficient (DSC) measures volumetric overlap between segmentation results and annotations. It is computed by:
\begin{equation}
    DSC(A,B)=\frac{2|A \cap B|}{|A|+|B|} ,
\end{equation}
where A is the sets of foreground voxels in the annotation and B is the corresponding sets of foreground voxels in the segmentation result, respectively.

Normalized surface distance (NSD)\cite{NSD} serves as a distance-based measurement to assess performance.  It is computed by:
\begin{equation}
    NSD(A, B)=\frac{\left|\partial A \cap R_{\partial B}^{(\tau)}\right|+\left|\partial B \cap R_{\partial A}^{(\tau)}\right|}{|\partial A|+|\partial B|} ,
\end{equation}
where $R_{\partial A}^{(\tau)}$ and $R_{\partial B}^{(\tau)}$ represent the borders of ground-truth and segmentation masks which use a threshold $\tau$ to tolerate the inter-rater variability of the annotators. We set $\tau= 3 mm $ for COVID-19 infection. In contrast to the DSC, which measures the overlap of volumes, the NSD measures the overlap of two surfaces.

We also consider three  other evaluation metrics. Accuracy denotes the correct rate for both positive and negative predictions. 
Sensitivity shows the percentage of positive instances correctly identified positive. F1-score is the weighted harmonic average of Precision and Sensitivity which is an effective and comprehensive evaluation.

\subsection{Analysis of Different Non-COVID19 Pre-trained Models}
\begin{figure*}[width=2.\linewidth,pos=htbp]
\flushleft
\includegraphics[scale=0.5]{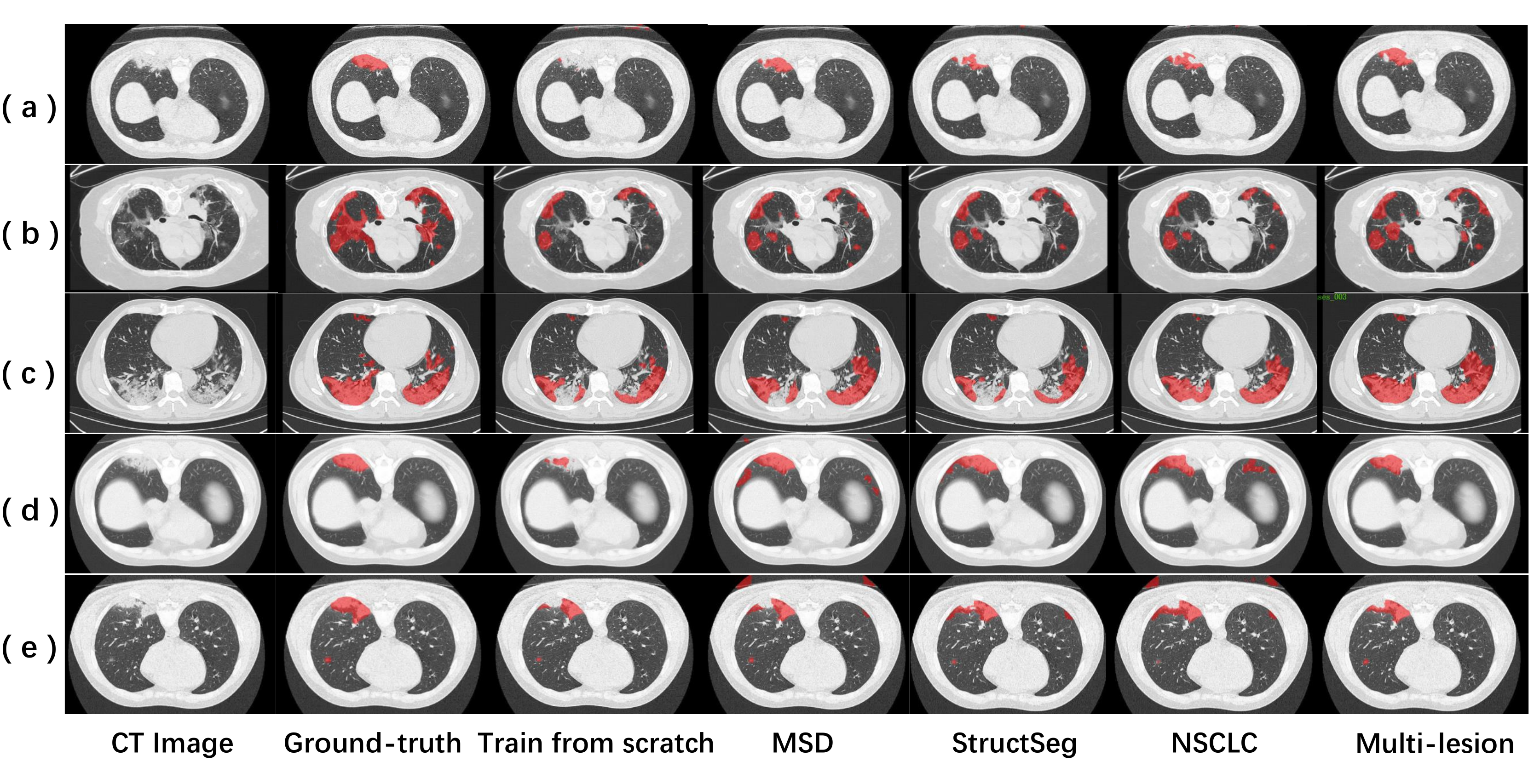}
\caption{Visual comparison of COVID-19 infection segmentation results. Training from scratch means using no pre-trained models. MSD, StructSeg, NSCLC and Multi-lesion mean using pre-trained models from corresponding non-COVID19 lesion datasets.}
\label{fig:pic2}

\end{figure*}
All of the four pre-trained models (MSD, StructSeg, NSCLC, Multi-lesion) are utilized to investigate the ability of transfer learning to COVID-19 datasets. With extensive experiments in Table \ref{1}, we observe that different pre-trained models show different transferability. The best scores of models corresponding to each transfer learning strategies are highlighted along with the detailed comparative analysis of 5 validation folds. 
\subsubsection{Single-lesion Pre-trained Model}Among pre-trained models on each single lesion (MSD, StructSeg, NSCLC), compared with training COVID-19 datasets from scratch, MSD tumor pre-trained model improves the segmentation by 3.0$\%$ DSC and 3.2$\%$ NSD at most. Meanwhile, StructSeg and NSCLC lung lesion pre-trained models show instability under different transfer strategies. As shown in Table \ref{1}, with Continual Learning and Body Fine-tuning strategies, StructSeg and NSCLC pre-trained models can achieve promising improvement in most folds, but instead, they obtain lower performance with the strategy of Pre-trained Lesion Representations. The rationale is that there exists a large domain distance between these two lesions and COVID-19 infection. The strategy of Pre-trained Lesion Representations totally uses encoded non-COVID19 lesion features. Thus, the transferability of a single-lesion dataset model largely depends on their domain difference. 
\begin{table*}[width=2.\linewidth,pos=htbp]
\renewcommand{\arraystretch}{1.3}
\caption{Results of 5-fold cross validation of different transfer learning strategies under different non-COVID19 lung lesion pre-trained models. The best results are shown in \textcolor{red}{\textbf{red}} font.}\label{1}
\centering{
\resizebox{\textwidth}{30mm}{
\begin{tabular}{c|c|cc|cc|cc|cc|cc|cc}
\toprule
\multirow{2}*{Transfer Strategy} &\multirow{2}{*}{\begin{tabular}[c]{@{}c@{}}Pre-trained \\model\end{tabular}}
     & \multicolumn{2}{c|}{fold 0} & \multicolumn{2}{c|}{fold 1} & \multicolumn{2}{c|}{fold 2} & \multicolumn{2}{c|}{fold 3} & \multicolumn{2}{c|}{fold 4} &\multicolumn{2}{c}{AVG}  \\ 
\cline{3-14}
     & & DSC & NSD & DSC & NSD  & DSC & NSD  & DSC & NSD  & DSC & NSD & DSC & NSD \\ 
\hline
\multirow{2}{*}{\begin{tabular}[c]{@{}c@{}}Train from \\scratch\end{tabular}} \
&\multirow{2}{*}{None} \ & \multirow{2}*{0.681$\pm$0.205}\  & \multirow{2}{*}{0.709$\pm$0.213} \  & \multirow{2}*{0.713$\pm$0.205} \ & \multirow{2}*{0.718$\pm$0.230}\  & \multirow{2}*{0.662$\pm$0.217} \  & \multirow{2}*{0.717$\pm$0.242} \  & \multirow{2}*{0.681$\pm$0.231} \ & \multirow{2}*{0.708$\pm$0.271}\  & \multirow{2}*{0.627$\pm$0.269} \  & \multirow{2}*{0.649$\pm$0.282} \  & \multirow{2}*{0.673$\pm$0.223} \  & \multirow{2}*{0.700$\pm$0.224} \\ &\ & \ & \ &&&&&&&&\\
\hline
\hline
\multirow{4}{*}{\begin{tabular}[c]{@{}c@{}}Continual \\Learning\end{tabular}}
& MSD  & 0.696$\pm$0.211\  & 0.725$\pm$0.219 \  & 0.731$\pm$0.206 \ & 0.734$\pm$0.238\  & \textbf{0.724$\pm$0.166} \  & \textbf{0.782$\pm$0.180} \  & 0.720$\pm$0.205 \ & 0.745$\pm$0.224\  & \textcolor{red}{\textbf{0.646$\pm$0.235}} \  & \textcolor{red}{\textbf{0.672$\pm$0.242}} \  & \textbf{0.703$\pm$0.203} \  & \textbf{0.732$\pm$0.219} \\

& StructSeg  & 0.672$\pm$0.234\  & 0.704$\pm$0.231\  & 0.712$\pm$0.206 	\  & 0.716$\pm$0.234 \  & 0.704$\pm$0.163 \  & 	0.762$\pm$0.183 \  & 	0.717$\pm$0.192 	\  & 0.745$\pm$0.208 \  & 	0.600$\pm$0.264 \  & 	0.616$\pm$0.276 \  & 	0.681$\pm$0.214\  &  	0.709$\pm$0.228  \\ 

& NSCLC  & 0.673$\pm$0.215 \  &	0.702$\pm$0.219 \  &	0.712$\pm$0.209 \  &	0.720$\pm$0.244 \  &	0.722$\pm$0.145 \  &	0.775$\pm$0.169 \  &	0.720$\pm$0.209 \  &	0.745$\pm$0.235 \  &	0.581$\pm$0.283 \  &	0.605$\pm$0.292 \  &	0.681$\pm$0.218 \  &	0.709$\pm$0.237  \\

& Multi-lesion  & \textbf{0.709$\pm$0.209} \  &	\textbf{0.726$\pm$0.220} \  &	\textbf{0.740$\pm$0.209} \  &	\textbf{0.735$\pm$0.232} \  &	0.692$\pm$0.191 \  &	0.745$\pm$0.198 \  &	\textbf{0.721$\pm$0.208} \  &	\textbf{0.749$\pm$0.219} \  &	0.618$\pm$0.251 \  &	0.638$\pm$0.260 \  &	0.696$\pm$0.213 \  &	0.718$\pm$0.225  \\
\hline
\hline
\multirow{4}{*}{\begin{tabular}[c]{@{}c@{}}Body \\Fine-tuning\end{tabular}}
& MSD  & 0.679$\pm$0.214 \  &	0.706$\pm$0.224 \  &	0.706$\pm$0.210 \  &	0.708$\pm$0.250 \  &	\textbf{0.724$\pm$0.157} \  &	\textbf{0.785$\pm$0.167} \  &	0.708$\pm$0.178 \  &	0.724$\pm$0.228 \  &	\textbf{0.623$\pm$0.247} \  &	\textbf{0.642$\pm$0.253} \  &	0.688$\pm$0.201 \  &	0.713$\pm$0.225 \\
& StructSeg  &0.689$\pm$0.203 \  &	0.710$\pm$0.222 \  &	0.719$\pm$0.207 \  &	0.721$\pm$0.238\  & 	0.713$\pm$0.161 \  &	0.770$\pm$0.173 \  &	0.724$\pm$0.209 \  &	0.743$\pm$0.231 \  &	0.603$\pm$0.265 \  &	0.617$\pm$0.271 \  &	0.689$\pm$0.211 \  &	0.712$\pm$0.229 \\
& NSCLC &0.696$\pm$0.197 \  &	0.714$\pm$0.213 \  &	0.716$\pm$0.206 \  &	0.720$\pm$0.239 \  &	0.673$\pm$0.208 \  &	0.734$\pm$0.219 \  &	0.690$\pm$0.229 \  &	0.707$\pm$0.266 \  &	0.579$\pm$0.288 \  &	0.590$\pm$0.298 \  &	0.671$\pm$0.228 \  &	0.693$\pm$0.248 \\
& Multi-lesion  & \textbf{0.702$\pm$0.209} \  &	\textbf{0.715$\pm$0.221} \  &	\textbf{0.736$\pm$0.209} \  &	\textbf{0.730$\pm$0.235}\  & 	0.703$\pm$0.182 	\  & 0.755$\pm$0.188\  & 	\textbf{0.725$\pm$0.211} \  &	\textbf{0.754$\pm$0.225}\  & 	0.612$\pm$0.254 \  &	0.628$\pm$0.263\  & 	\textbf{0.696$\pm$0.213} \  & 		\textbf{0.717$\pm$0.227} \\
\hline
\hline
\multirow{4}{*}{\begin{tabular}[c]{@{}c@{}}Pre-trained \\Lesion\\Representations\end{tabular}}
& MSD  & \textbf{0.709$\pm$0.210} \  & 	\textbf{0.735$\pm$0.225} \  & 	\textcolor{red}{\textbf{0.741$\pm$0.207}} \  & 	\textcolor{red}{\textbf{0.751$\pm$0.233}} \  & 	\textbf{0.713$\pm$0.159} \  & 	\textbf{0.763$\pm$0.173} \  & 	0.693$\pm$0.217 \  & 		0.689$\pm$0.262 \  & 	\textbf{0.618$\pm$0.249} \  & 	\textbf{0.637$\pm$0.258} \  & 	\textbf{0.695$\pm$0.209} \  & 	0.715$\pm$0.232\\ 
& StructSeg  &0.657$\pm$0.231 \  & 	0.670$\pm$0.243 \  & 	0.696$\pm$0.215 \  & 	0.687$\pm$0.249 \  & 	0.657$\pm$0.233 \  & 	0.701$\pm$0.246 \  & 	0.682$\pm$0.231 \  & 	0.701$\pm$0.237 \  & 	0.536$\pm$0.271 \  & 	0.535$\pm$0.276 \  & 	0.645$\pm$0.238 \  & 	0.659$\pm$0.252 \\
& NSCLC  & 0.663$\pm$0.214 \  & 	0.668$\pm$0.224 \  & 	0.695$\pm$0.209 \  & 	0.683$\pm$0.241 \  & 	0.626$\pm$0.240 \  & 	0.664$\pm$0.251 \  & 	0.658$\pm$0.229 \  & 	0.670$\pm$0.254 \  & 	0.544$\pm$0.285 \  & 	0.543$\pm$0.294 \  & 	0.637$\pm$0.237 \  & 	0.646$\pm$0.253 \\
& Multi-lesion & 0.704$\pm$0.210 \  & 	0.728$\pm$0.220 \  & 	0.731$\pm$0.208 \  & 	0.737$\pm$0.235 \  & 	0.705$\pm$0.193 \  & 	0.748$\pm$0.213 \  & 	\textbf{0.725$\pm$0.209} \  & \textbf{0.751$\pm$0.221} \  & 	0.602$\pm$0.265 \  & 	0.618$\pm$0.271 \  & 	0.693$\pm$0.218 \  & 		\textbf{0.716$\pm$0.232} \\
\hline
\hline
\multirow{1}*{Hybrid-encoder} \ &
Multi-lesion  & \textcolor{red}{\textbf{0.712$\pm$0.207}}\  &	\textcolor{red}{\textbf{0.737$\pm$0.225}}\  &\textbf{0.731$\pm$0.205}\  &	\textbf{0.745$\pm$0.237}\  &	\textcolor{red}{\textbf{0.726$\pm$0.132}}\  &	\textcolor{red}{\textbf{0.783$\pm$0.163}}\  &	\textcolor{red}{\textbf{0.736$\pm$0.131}}\  &	\textcolor{red}{\textbf{0.778$\pm$0.143}}\  &	\textbf{0.615$\pm$0.254}\  &	\textbf{0.634$\pm$0.269}\  &	\textcolor{red}{\textbf{0.704$\pm$0.206}}\  &	\textcolor{red}{\textbf{0.735$\pm$0.224}}\\
\bottomrule

\end{tabular}}}
\end{table*}
\subsubsection{Multi-lesion Pre-trained Model} As shown in Table \ref{1}, we further notice that among all transfer learning strategies, multi-lesion pre-trained model not only achieves a high percentage for DSC and NSD values, but also performs the most stably among all pre-trained models, with the average DSC value of 0.696, 0.696, 0.693, 0.704, respectively. This shows the robustness of a multi-lesion pre-trained model used for transfer learning to COVID-19 infection. Table \ref{3} further verifies this demonstration using Sensitivity, F1-score and Accuracy concerned with different transfer strategies. It is observed that transferring from multi-lesion pre-trained model outperforms training from scratch on all these evaluation metrics among all strategies. 
Significantly, when it comes to Pre-trained Lesion Representations, where the encoder of pre-trained models is totally frozen and serves as a non-COVID19 lesion features extractor, the multi-lesion pre-trained model can still perform well. This confirms the effectiveness of multi-task training for multiple lung lesions, which generates more robust and general-purpose representations to help COVID-19 infection tasks. Fig.\ref{fig:pic2}(a)-(e) show some examples of segmentation results of the above pre-trained models. It is clear that compared with training from scratch, pre-training from single non-COVID19 models can obtain more accurate massive structures of COVID-19 but shows dissatisfying instability. However, pre-training from multi-lesion model shows high precision and smooth boundary like manually annotated.
\begin{table}[htbp]  \caption{Results of Average Sensitivity, F1-score and Accuracy of different transfer learning strategies under different non-COVID19 lung lesion pre-trianed models. The best results are shown in \textcolor{red}{\textbf{red}} font.}
\renewcommand{\arraystretch}{1.}
 \label{3}

 \resizebox{.5\textwidth}{38mm}{\begin{tabular}{cccc}

\toprule   

   & Sensitivity & F1-score & Accuracy\\  
\midrule
Train from scratch & 0.6204$\pm$0.2369\  & 	0.6728$\pm$0.2227 \  & 	0.9939$\pm$0.0086    \\

\midrule  
\midrule
\multicolumn{2}{l}{Continual Learning}\\
\midrule
MSD & \textbf{0.6592$\pm$0.2148} \  & 	\textbf{0.7033$\pm$0.2029} \  & 	\textbf{0.9942$\pm$0.0083 }   \\
StructSeg &  0.6317$\pm$0.2224 \  & 	0.6809$\pm$0.2139 \  & 	0.9941$\pm$0.0081    \\    
NSCLC &0.6371$\pm$0.2249 \  & 	0.6814$\pm$0.2180 \  & 	0.9939$\pm$0.0086     \\
Multi-lesion &0.6545$\pm$0.2244 \  & 	0.6962$\pm$0.2134 \  & 	0.9941$\pm$0.0083     \\
\midrule
\midrule
\multicolumn{2}{l}{Body Fine-tuning}\\
\midrule
MSD & 0.6482$\pm$0.2186 \  & 	0.6877$\pm$0.2170 \  & 	0.9934$\pm$0.0100    \\
StructSeg &  0.6392$\pm$0.2312 \  & 	0.6894$\pm$0.2112\  &  	0.9940$\pm$0.0083    \\    
NSCLC &0.6270$\pm$0.2041\  &  	0.6709$\pm$0.2277 \  & 	0.9941$\pm$0.0086    \\
Multi-lesion & \textbf{0.6528$\pm$0.2252 }\  & 	\textbf{0.6956$\pm$0.1328} \  & \textbf{0.9941$\pm$0.0083}     \\
\midrule
\midrule
\multicolumn{2}{l}{Pre-trained Lesion Representations} \\
\midrule
MSD & \textbf{0.6809$\pm$0.2055} \  & 	\textbf{0.6949$\pm$0.2093} \  & 0.9940$\pm$0.0083   \\
StructSeg &  0.6218$\pm$0.2362 \  & 	0.6455$\pm$0.2382 \  & 	0.9926$\pm$0.0118    \\    
NSCLC &0.6200$\pm$0.2376 \  & 	0.6371$\pm$0.2368 \  & 	0.9929$\pm$0.0096     \\
Multi-lesion &0.6622$\pm$0.2117 \  & 0.6934$\pm$0.2185\  & \textbf{0.9940$\pm$0.0082}    \\
\midrule
\midrule
\multicolumn{2}{l}{Hybrid-encoder}\\
\midrule
Multi-lesion &\textcolor{red}{\textbf{0.6818$\pm$0.2147}} \  & \textcolor{red}{\textbf{0.7069$\pm$0.1870} }\  & \textcolor{red}{\textbf{0.9943$\pm$0.0081}}\\
\bottomrule  

\end{tabular}}
\end{table}

\subsection{Analysis of Different Transfer Learning Strategies}
Based on the above conclusion that multi-lesion pre-trained model brings consistently higher accuracy and robustness, we further analyze the performance of different transfer strategies adopted in this paper.

Table \ref{2} shows the results of COVID-19 infection segmentation using the same multi-lesion pre-trained model under different transfer strategies. These results suggest that all the transfer strategies improve the performance of training from scratch on average DSC, NSD and Sensitivity. In particular, in fold 2, they improve the segmentation by more than 6.4$\%$ DSC and 6.6$\%$ NSD on maximum. 
The strategies of Continual Learning and Body Fine-tuning get similar promising results, which both improve the segmentation by 2.3$\%$ DSC, 1.8$\%$ NSD and 3.4$\%$ Sensitivity on average. 

An interesting observation is that, compared with Continual Learning and Body Fine-tuning, where all the parameters are updated on COVID-19 dataset, the strategy of Pre-trained Lesion Representations still achieves a competitive performance with an entirely frozen encoder and inherited weights of pre-trained non-COVID19 models. It is observed to improve the DSC from 0.673 to 0.693, NSD from 0.700 to 0.716 and Sensitivity from 0.643 to 0.662. In particular, in terms of training cost, Table \ref{2} shows the training time (per epoch) for each transfer strategy. It can be clearly seen that the strategy of Pre-trained Lesion Representations spends just 148s per epoch on average, much less than training from scratch and other transfer strategies. Due to the frozen encoder, the strategy of Pre-trained Lesion Representations cuts down nearly a half of parameters that need to be updated. Thus, it is promising to adopt this strategy for COVID-19 transfer learning to save training costs and gain fast convergence.

\begin{figure*}[width=2.\linewidth,pos=htbp]
\flushleft
\includegraphics[scale=0.5]{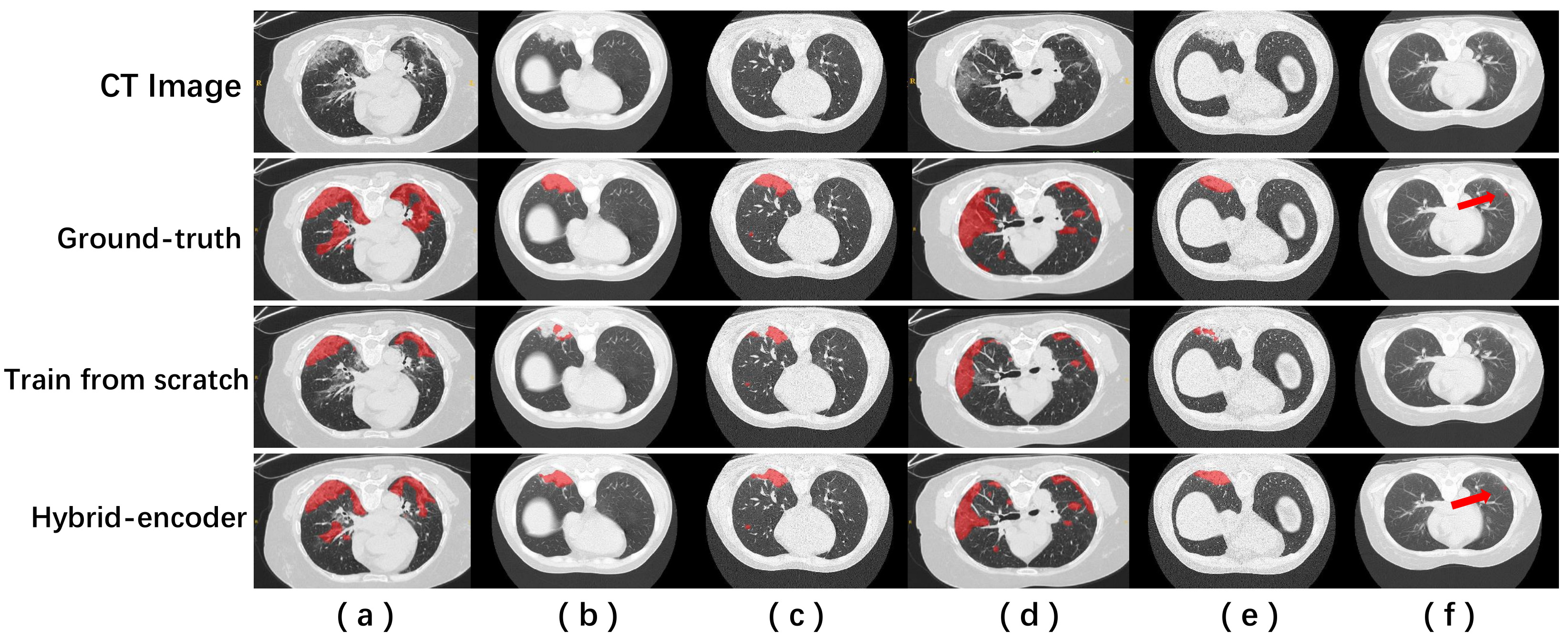}
\caption{Visual comparison of COVID-19 infection segmentation results between training from scratch and proposed Hybrid-encoder transfer learning strategy based on multi-lesion pre-trained model.}
\label{fig:pic1}
\end{figure*}

It is also observed in Table \ref{2} that our proposed Hybrid-encoder transfer learning strategy exhibits significantly better segmentation performance than other strategies using multi-lesion pre-trained model. It improves the average DSC from 0.673 to 0.704 and NSD from 0.700 to 0.735, which also performs best among all the pre-trained models in Table \ref{1}. In terms of Sensitivity, F1-score and Accuracy in Table \ref{3}, proposed Hybrid-encoder learning outperforms other strategies and enhances the values to 0.6818, 0.7069 and 0.9943, respectively. 
The transfer ability of Hybrid-encoder is also confirmed by Fig.\ref{fig:pic1}. Compared with training from scratch, Hybrid-encoder learning yields segmentation results with more accurate boundaries in Fig.\ref{fig:pic1} (b)(c)(e) and identifies some minor COVID-19 infection areas in Fig.\ref{fig:pic1} (a)(d)(f). 
The success of proposed Hybrid-encoder learning strategy is owed to the designed parallel encoders, where the COVID-19 and non-COVID19 lesions are both employed for encoding feature representations, leading to better generalization and lower over-fitting risks.

\begin{table*}[width=2.5\linewidth,pos=!htbp]
\renewcommand{\arraystretch}{1.3}
\renewcommand{\multirowsetup}{\centering}
\caption{Results of 5-fold cross validation of different transfer learning strategies based on Multi-lesion pre-trained model. \\
\textbf{Time} represents the training time for an epoch. The best results are shown in \textcolor{red}{\textbf{red}} font.}\label{2}
\centering{
\resizebox{\textwidth}{28mm}{\begin{tabular}{c|c|cccc|cccc|cccc|cccc|cccc}
\toprule
     &&\multicolumn{4}{c|}{Train from scratch} & \multicolumn{4}{c|}{Continual Learning} & \multicolumn{4}{c|}{Body Fine-tuning} 
     & \multicolumn {4}{c|}{Lesion Representations}
     & \multicolumn {4}{c}{Hybrid-encoder}\\ 
\cline{3-22}
     && DSC$\uparrow$ & NSD$\uparrow$ & Sen.$\uparrow$& Time$\downarrow$ & DSC$\uparrow$ & NSD$\uparrow$ & Sen.$\uparrow$& Time$\downarrow$  & DSC$\uparrow$ & NSD$\uparrow$ & Sen.$\uparrow$& Time$\downarrow$ & DSC$\uparrow$ & NSD$\uparrow$ & Sen.$\uparrow$& Time$\downarrow$& DSC$\uparrow$ & NSD$\uparrow$ & Sen.$\uparrow$& Time$\downarrow$\\
\hline
\multirow{2}{*}{fold 0}
&avg& 0.681\  & 0.709\  & 0.639\  & 224s \ & 0.709\  & 0.726 \  & 0.683 \  & 194s \ & 0.702\  & 0.715 \  & 0.668 \  & 196s\  & 0.704\  & 0.728 \ & 0.682\  & \textcolor{red}{\textbf{145s}}\ & \textcolor{red}{\textbf{0.712}}\ & \textcolor{red}{\textbf{0.737}}\ & \textcolor{red}{\textbf{0.696}}\ &  212s\\
&std& 0.205\  & 0.213\  & 0.162\  &\ \ --  \ & 0.209\  & 0.220 \  & 0.161 \  & \ \ -- \ & 0.209\  & 0.221 \  & 0.160 \  &\ \ -- \  & 0.210\  & 0.220 \ & 0.156\  &\ \ -- \ & 0.207\ & 0.225\ & 0.153\ & \ \ --\\
\hline
\multirow{2}{*}{fold 1}  
&avg& 0.713\  & 0.718\  & 0.707\  & 194s \ & \textcolor{red}{\textbf{0.740}}\  & 0.735 \  & 0.752 \  & 193s \ & 0.736\  & 0.730 \  & 0.754 \  & 196s\  & 0.731\  & 0.737 \ & 0.762\  & \textcolor{red}{\textbf{145s}}\ & 0.731\ & \textcolor{red}{\textbf{0.745}}\ & \textcolor{red}{\textbf{0.773}}\ & 210s\\
&std& 0.205\  & 0.230\  & 0.189\  & \ \ -- \ & 0.209\  & 0.232 \  & 0.204 \  & \ \ -- \ & 0.209\  & 0.235 \  & 0.172 \  & \ \ --\  & 0.208\  & 0.235 \ & 0.126\  & \ \ --\ & 0.205\ & 0.237\ & 0.132\ &  \ \ --\\
\hline
\multirow{2}{*}{fold 2}  
&avg& 0.662\  & 0.717\  & 0.621\  & 224s \ & 0.692\  & 0.745 \  & 0.667 \  & 193s \ & 0.703\  & 0.755 \  & 0.688 \  & 194s\  & 0.705\  & 0.748 \ & 0.676\  & \textcolor{red}{\textbf{146s}}\ & \textcolor{red}{\textbf{0.726}}\ & \textcolor{red}{\textbf{0.783}}\ & \textcolor{red}{\textbf{0.701}}\ & 212s\\ 
&std& 0.217\  & 0.242\  & 0.295\  &  \ \ -- \ & 0.191\  & 0.198 \  & 0.272 \  &  \ \ -- \ & 0.182\  & 0.188 \  & 0.262 \  &  \ \ --\  & 0.193\  & 0.213 \ & 0.260\  &  \ \ --\ & 0.123\ & 0.163\ & 0.240\ &   \ \ --\\
\hline
\multirow{2}{*}{fold 3}  
&avg & 0.681\  & 0.708\  & 0.604\  & 194s \ & 0.721\  & 0.749 \  & 0.657 \  & 193s \ & 0.725\  & 0.754 \  & 0.648 \  & 196s\  & 0.725\  & 0.751 \ & 0.696\  & \textcolor{red}{\textbf{148s}}\ & \textcolor{red}{\textbf{0.736}}\ & \textcolor{red}{\textbf{0.778}}\ & \textcolor{red}{\textbf{0.705}}\ &  215s\\
&std& 0.231\  & 0.271\  & 0.241\  &  \ \ -- \ & 0.208\  & 0.219 \  & 0.164 \  &  \ \ -- \ & 0.211\  & 0.225 \  & 0.203 \  &  \ \ --\  & 0.209\  & 0.221 \ & 0.143\  &  \ \ --\ & 0.131\ & 0.143\ & 0.147\ &   \ \ --\\
\hline
\multirow{2}{*}{fold 4}  
&avg & \textcolor{red}{\textbf{0.627}}\  & \textcolor{red}{\textbf{0.649}}\  & 0.531\  & 194s \ & 0.618\  & 0.638 \  & 0.514 \  & 223s \ & 0.612\  & 0.628 \  & 0.506 \  & 194s\  & 0.602\  & 0.618 \ & 0.495\  & \textcolor{red}{\textbf{157s}}\ & 0.615\ & 0.634\ & \textcolor{red}{\textbf{0.534}}\ &  218s\\ 
&std& 0.269\  & 0.282\  & 0.267\  &  \ \ -- \ & 0.251\  & 0.260 \  & 0.254 \  &  \ \ -- \ & 0.254\  & 0.263 \  & 0.258 \  & 196s\  & 0.265\  & 0.271 \ & 0.255\  &  \ \ --\ & 0.254\ & 0.269\ & 0.263\ &   \ \ --\\
\hline
\multirow{2}{*}{AVG}  
&avg & 0.673\  & 0.700\  & 0.620\  & 206s \ & \textbf{0.696}\  & \textbf{0.718} \  & \textbf{0.655} \  & \textbf{199s} \ & \textbf{0.696}\  & \textbf{0.717} \  & \textbf{0.653} \  & \textbf{195s}\  & \textbf{0.693}\  & \textbf{0.716} \ & \textbf{0.662}\  & \textcolor{red}{\textbf{148s}}\ & \textcolor{red}{\textbf{0.704}}\ & \textcolor{red}{\textbf{0.735}}\ & \textcolor{red}{\textbf{0.682}}\ &  \textbf{213s}\\
&std& 0.223\  & 0.224\  & 0.237\  & 16s \ & 0.213\  & 0.225 \  & 0.224 \  & 13s \ & 0.213\  & 0.227 \  & 0.225 \  & \ 1s\  & 0.218\  & 0.232 \ & 0.212\  & \ 5s\ & 0.206\ & 0.224\ & 0.200\ & \  3s\\

\bottomrule
\end{tabular}}}
\end{table*}

\section{Discussion}
A valuable demonstration is that a multi-lesion pre-trained model can make the best of multiple lesion representations, and advance the generalization and robustness of pre-trained models. There is value in recognizing that the CT appearance of these different lung lesions share some similarity. Thus, with more kinds of lung lesion datasets incorporated to pre-train a model, we could achieve better performance. This exploration is an important contribution, enabling more research on transfer learning to COVID-19 infection from the perspective of utilizing non-COVID19 lung lesions. 

Moreover, this paper examines a series of different transfer learning strategies, including Continual Learning, Body Fine-tuning, Pre-trained Lesion Representations and the proposed Hybrid-encoder Learning. We observe segmentation improvement in all performance metrics. It is also noted that the strategy of Pre-trained Lesion Representations with a frozen encoder enhances performance as well. This gains more insight into the significant transferability exhibited by the encoding parameters. Though the encoding layers do not contain any explicit knowledge of the COVID-19 infection, their parameters still enable the optimizer to reach a higher performance while fine-tuning.
The rationale is that the encoded multi-lesion representations contain more high-level and abundant encoding information of the medically relevant lung lesion observed in CT images. Through combining the multi-lesion representations and COVID-19 infection features, the proposed Hybrid-encoder achieves significant improvement. 
These observation and exploration are important not only in COVID-19 transfer learning but also in the general medical domain, because feature reuse from pre-training out-of-domain datasets shows significant improvement for task performance and training convergence.
\section{Conclusion}
In this paper, we investigate the transferability in COVID-19 CT segmentation. We present a set of experiments to better understand how different non-COVID19 lung lesions influence the performance of COVID-19 infection segmentation and their different transfer ability under different transfer learning strategies. Our results reveal clear benefits of pre-training on non-COVID19 lung lesion datasets when public labelled COVID-19 datasets are inadequate to train a robust deep learning model. Among all the strategies, our proposed Hybrid-encoder Learning method based on multi-lesion pre-trained model effectively utilizes tranferred non-COVID19 lung lesion knowledge and gains significant improvement.

Future research directions include utilizing more various non-COVID19 lung lesion datasets and investigating better transfer learning methods, so that non-COVID19 lung lesions can be effectively used to improve the quality of COVID-19 infection segmentation in the absence of sufficient high-quality COVID-19 datasets. 
\section*{Declaration of Competing Interest}
The authors have no conflict of interest to disclose. 
\bibliographystyle{IEEEtran}
\bibliography{reference}




%








\end{document}